\documentclass[aps,prd,onecolumn,groupedaddress]{revtex4-1}

\usepackage{graphicx}
\usepackage{booktabs}
\usepackage{amssymb,bm,mathrsfs,bbm,amscd}
\usepackage[tbtags]{amsmath}
\usepackage{lineno}

\begin{document}

\title{Testing Lorentz symmetry with space-based gravitational-wave detectors}
\author{Cheng-Gang Qin$^1$, Jun Ke$^{2}$, Qin Li$^1$, Ya-Fen Chen$^3$}
\author{Jie Luo$^{2}$} \email[E-mail:]{luojiethanks@126.com}
\author{Yu-Jie Tan$^{1}$} \email[E-mail:]{yjtan@hust.edu.cn}
\author{Cheng-Gang Shao$^1$}\email[E-mail:]{cgshao@hust.edu.cn}

\affiliation
{$^1$MOE Key Laboratory of Fundamental Physical Quantities
Measurement $\&$ Hubei Key Laboratory of Gravitation and Quantum Physics, PGMF and School of Physics, Huazhong University of Science and Technology,
Wuhan 430074,  People's Republic of China\\
$^2$ School of Mechanical Engineering and Electronic Information, China University of Geosciences, Wuhan 430074, People's Republic of China\\
$^3$ School of Mathematics and Physics, Hubei Polytechnic University, Huangshi 435003, People's Republic of China}

%\date{\today}

\begin{abstract}
Lorentz symmetry (LS), one of the most fundamental physical symmetries, has been extensively studied in the context of quantum gravity and unification theories. Many of these theories predict a LS violation, which could arise from the discreteness of spacetime, or extra dimensions.
Standard-model extension (SME) is an effective field theory to describe Lorentz violation whose effects can be explored using precision instruments such as atomic clocks and gravitational-wave (GW) detectors. Considering the pure-gravity sector and matter-gravity coupling sector in the SME, we studied the leading Lorentz-violating modifications to the time delay of light and the relativistic frequency shift of the clock in the space-based GW detectors.
We found that the six data streams from the GW mission can construct various combinations of measurement signals, such as single-arm round-trip path, interference path, triangular round-trip path, etc.
These measurements are sensitive to the different combinations of SME coefficients and provide novel linear combinations of SME coefficients different from previous studies.
Based on the orbits of TianQin, LISA, and Taiji missions, we calculated the response of Lorentz-violating effects on the combinations of the measurement signal data streams.
Our results allow us to estimate the sensitivities for SME coefficients: $10^{-6}$ for the gravity sector coefficient $\bar{s}^{TT}$, $10^{-6}$ for matter-gravity coupling coefficients $(\bar{a}^{(e+p)}_{\text{eff}})_{T}$ and $\bar{c}^{(e+p)}_{TT}$, and $10^{-5}$ for $(\bar{a}^{n}_{\text{eff}})_{T}$ and $\bar{c}^{n}_{TT}$.
\end{abstract}

\maketitle

\section{introduction}\label{sct1}

Lorentz symmetry plays a crucial role in both the General Relativity (GR) and Standard Model (SM) of particle physics. General Relativity brilliantly describes the gravitational phenomena at the macroscopic scales, which has survived every experimental test \cite{will2014confrontation,will2018theory}. On the other hand, the Standard Model of particle physics provides a highly successful interpretation of matter and nongravitational interactions in the quantum domain. These two theories are the cornerstones of our current understanding of the natural world. However, it remains a challenging task to merge the GR with the SM into a single underlying unified theory. Lorentz symmetry, as a fundamental symmetry in both theories, may be the key to this challenge. Many theories, attempting to unify a microscopic description via quantum physics with the macroscopic and classical description of gravitational effects, predict a tiny violation of Lorentz symmetry at the Planck scale, including string theory \cite{kostelecky1989spontaneous}, noncommutative theories \cite{PhysRevLett.87.141601}, loop quantum gravity \cite{PhysRevD.65.103509}, supersymmetry \cite{PhysRevLett.94.081601}, etc. Experimental probing of these predictions at such high energy level are currently infeasible; however, at low energy scales, the suppressed new physics effects arising from Lorentz violations can be detected with present experiments \cite{Mattingly2005Modern,Tasson2014What}.

An effective field theory that describes the observable effects of Lorentz violation in an underlying theory of quantum gravity is the Standard-Model Extension (SME). The SME is constructed by adding all possible Lorentz-violating terms in all sectors of physics \cite{PhysRevD.51.3923,PhysRevD.69.105009,PhysRevD.58.116002}. It contains both general relativity and the standard model Lagrange densities, along with possible Lorentz-violating terms from the gravitational and standard-model fields. This theoretical framework provides a powerful approach to studying the experimental signals of Lorentz violations. By specifying the SME coefficient values, numerous experimental predictions have been proposed with attainable sensitivities in terms of SME \cite{PhysRevD.58.116002,PhysRevD.66.056005,PhysRevD.74.045001,PhysRevD.80.015020,PhysRevD.83.016013,PhysRevD.88.105011,PhysRevD.93.025020,PhysRevD.95.075026,sanner2019optical,PhysRevD.101.064056,PhysRevD.104.044054}.
The matter-sector tests of the minimal SME have been studied in various experiments, which includes the photon sector \cite{PhysRevLett.90.060403,PhysRevLett.90.211601,PhysRevD.71.025004,PhysRevD.87.125028,PhysRevD.89.105019,
PhysRevLett.91.020401,PhysRevD.97.115043,PhysRevD.100.055036}, electron sector \cite{PhysRevLett.84.1381,PhysRevLett.90.201101,PhysRevD.68.116006,PhysRevD.70.076004,PhysRevD.71.045004,gomes2022constraining}, and nucleon sector \cite{PhysRevLett.86.3228,PhysRevLett.96.060801,PhysRevLett.111.151102,PhysRevD.64.076001,PhysRevD.72.087702}. On the other hand, the gravity and matter-gravity coupling sectors in post-Newtonian gravity have been investigated in experiments such as the lunar and satellite laser ranging \cite{PhysRevLett.99.241103,PhysRevD.103.064055,PhysRevLett.117.241301,PhysRevLett.119.201102,PhysRevD.103.064055}, short-range gravity experiments \cite{PhysRevLett.122.011102,PhysRevLett.117.071102,PhysRevD.91.022006,PhysRevD.91.092003,PhysRevD.91.102007,kostelecky2017testing}, laboratory measurements with gravimeters \cite{PhysRevLett.100.031101,PhysRevLett.119.201101,PhysRevD.80.016002,PhysRevD.97.024019}, the timing of the binary pulsar \cite{PhysRevD.98.084049,shao2012new,PhysRevD.92.125028,PhysRevLett.112.111103,PhysRevD.103.084028}, the time delay of light \cite{PhysRevD.80.044004,PhysRevD.105.124069} and clock-comparison experiments \cite{PhysRevLett.88.090801,PhysRevD.98.036003,qin2021test,PhysRevD.100.084022}. For a summary of the experimental constraints on SME coefficients, it can be found in Ref. \cite{RevModPhys.83.11}. To date, though no compelling evidence of Lorentz violations for non-zero coefficients has been found, an incredible number of opportunities still exist for further investigations, especially through experiments with precision instruments.

Given the successful detection of gravitational waves (GWs) by ground-based GW detectors LIGO and VIRGO \cite{PhysRevLett.116.061102,PhysRevLett.116.241103,PhysRevLett.118.221101,PhysRevLett.119.141101,PhysRevLett.119.161101,abbott2017gw170608}, space-based GW detectors have recently garnered significant attention. By using the high-precision instruments, GW detectors could achieve excellent sensitive levels at or beyond quantum-noise limits. Apart from GW physics, this also motivates GW detectors to probe the fundamental physics questions including dark matter \cite{vermeulen2021direct}, and violation of Lorentz symmetry \cite{kostelecky2016testing,kostelecky2016searching,ellis2019limits,xu2020neutron,PhysRevD.106.124019,schreck2017fermionic}. For instance, the GW observatory is associated with laser interferometry with a huge size, and it can be used to search Lorentz-violating effects \cite{PhysRevD.85.024041,PhysRevD.91.082003,PhysRevD.99.104062,PhysRevD.105.044034}. The preliminary data of LIGO has been used to search for the violations of Lorentz symmetry \cite{kostelecky2016searching}. Therefore, it is natural to ask about the potential of space-based GW observatories. For the space GW missions, there are two classic tests for Lorentz violations involving the gravitational time delays and frequency shift of clock comparison.
Improvements in two-way laser interferometer between satellites have led to significant improvements in displacement measurements, which provide accurate information about the light propagation in the gravitational field. The ultra-stable oscillators on satellites are used to provide the frequency reference, which provides measurements for the Doppler shift and gravitational redshift.
All these measurements are, in principle, sensitive to the specific combinations of SME coefficients. In space-based GW missions, it is therefore important to analyze the signals of Lorentz violation in detail.

In recent years, several space-based GW missions are underway or planned, such as LISA, \cite{amaro2017laser} TianQin, \cite{luo2016tianqin}, and Taiji \cite{hu2017taiji}. In this study, we focus on their potential sensitivity to the Lorentz-violating effects. These missions utilize three satellites to form an equilateral triangular configuration, which can provide six data streams. These six data streams can be combined in various ways to construct different data combinations, including single-arm round-trip path, interference path, and triangular round-trip paths. These data combinations are sensitive to different linear combinations of SME coefficients. The results may provide a way towards a full decorrelation of individual SME coefficients.

The paper is organized as follows: In Sec. \ref{sct2}, we briefly present the basic information of the pure-gravity and matter-gravity coupling sectors of SME theory, and we also introduce the orbit coordinate frame for the following analysis.
In Sec. \ref{sct3}, we study the Lorentz-violating effects of classic tests in the space-based GW missions, which include time delay of light, and frequency shift of clock. For these classic tests, we obtain the sensitive combinations of SME coefficients. Considering the GW mission TianQin, LISA, and Taiji, we estimate the attainable sensitivities for the SME coefficients.
In Sec. \ref{sct4}, based on the fundamental orbits of TianQin, LISA, and Taiji, and we evaluate and analyze the signals for general-relativity effects and Lorentz-violating effects.
Finally, we give the conclusion in Sec. \ref{sct5}.

\section{Theory frame of SME}\label{sct2}

The SME with gravitational coupling and matter-gravity coupling has been presented in Refs. \cite{PhysRevD.69.105009,PhysRevD.74.045001,PhysRevD.104.044054}. In the general scenario, the geometric spacetime framework is a Riemann-Cartan spacetime, which includes couplings to curvature and torsion degrees of freedom. In the limit of Riemann spacetime, the relevant action for pure-gravity and matter-gravity coupling sector of the SME can be written as
\begin{eqnarray}\label{tsme1}
  S  =\frac{1}{16\pi G}\int d^{4} x\sqrt{-\textsl{g}}[(1-u)R+  s^{\mu\nu}R^{T}_{\mu\nu}
  + t^{\kappa\lambda\mu\nu}C_{\kappa\lambda\mu\nu}]-mc\int d\lambda \left(\sqrt{-(g_{\mu\nu}+c_{\mu\nu})u^{\mu}u^{\nu}}+\frac{(a_{\text{eff}})_{\mu}}{m}u^{\mu}\right)+S',
\end{eqnarray}
where $\textsl{g}$ is the determinant of the spacetime metric $\textsl{g}_{\mu\nu}$, $R$ is the Ricci scalar, $R^{T}_{\mu\nu}$ is the trace-free Ricci tensor, $C_{\kappa\lambda\mu\nu}$ is the Weyl conformal tensor, $\lambda$ is the affine parameter, $u^{\mu}=d x^{\mu}/d \lambda$ is the four-velocity of the particle $m$. Note that in Eq.(\ref{tsme1}), the first part of the action is directly taken from the gravitational sector of the minimal SME, and the second part is derived from the SME via a complicated (nonlinear) map from the SME Dirac fermion sector to a classical-particle Lagrangian \cite{kostelecky2010classical}. The coefficients for Lorentz violation $u$, $s^{\mu\nu}$  and $t^{\kappa\lambda\mu\nu}$ describe the Lorentz-violating gravitational couplings. The coefficients $(a_{\text{eff}})_{\mu}$ and $c_{\mu\nu}$ are the Lorentz-violating fields for the matter sector. We neglect the contribution of $u$ and $t^{\kappa\lambda\mu\nu}$. In the post-Newtonian approximation, the $u$ can be eliminated by a redefinition of the metric tensor. $t^{\kappa\lambda\mu\nu}$ term can lead to some significant effects in cosmology \cite{PhysRevD.91.125002,PhysRevD.96.044036}, however, for the experiments considered in this paper, it does not affect the experimental results at the leading order. Then, we only focus on the influences of observable fields $s^{\mu\nu}$, $(a_{\text{eff}})_{\mu}$ and $c_{\mu\nu}$.

In the case of weak gravitational fields, such as the field of the Solar System, the signals of Lorentz violation can be studied in the post-Newtonian metric. For the Lorentz-violating terms, the pure-gravity sector is controlled by the coefficients $\bar s^{\mu\nu}$, and the relevant coefficients of the matter-gravity coupling sector are given by $(\bar{a}_{\text{eff}})_{\mu}$ and $\bar{c}_{\mu\nu}$. In harmonic coordinates, the metric is given by \cite{PhysRevD.74.045001,PhysRevD.84.085025}
\begin{eqnarray}\label{am}
   g_{00}&=&-1+\frac{1}{c^{2}}\left[ {\left(2+3\bar s_{00}+2\bar{c}^{s}_{00}+4\frac{\alpha}{M}\left(\bar{a}^{s}_{\text{eff}}\right)_{0}   \right)U}+{\bar s_{ij}U^{ij}} \right]+O(c^{-3}),\nonumber\\
  g_{0i}&=& \frac{1}{c^{2}} \left[ \left( \bar s_{0i}+\frac{\alpha}{M}\left(\bar{a}^{s}_{\text{eff}}\right)_{i}\right) U+\left(\bar s_{0k}+\frac{\alpha}{M}\left(\bar{a}^{s}_{\text{eff}}\right)_{k}\right) U^{ik} \right]+O(c^{-3}),\nonumber\\
  g_{ij}&=&\delta_{ij}+\frac{1}{c^{2}} \Big[ \left(2-\bar s_{00}+2\bar{c}^{s}_{00}-2\frac{\alpha}{M}\left(\bar{a}^{s}_{\text{eff}}\right)_{0} \right)\delta^{ij}U \nonumber\\
   &+&\left(\bar s_{lk}\delta^{ij}-\bar s_{lj}\delta^{ik}-\bar s_{ik}\delta^{lj}+2\left(\bar s_{00}+\frac{\alpha}{M}\left(\bar{a}^{s}_{\text{eff}}\right)_{0}\right)\delta^{il}\delta^{jk} \right)U^{lk} \Big]+O(c^{-3}),
\end{eqnarray}
where $M$ is the mass of gravitational source $s$. $g_{0i}$ components contain the contribution of SME coefficients to $c^{-2}$ terms and neglect the contributions by Lense-Thirring term, since the Lense-Thirring term is the order of $c^{-3}$ that is a higher-order effect for the following studies. The coupling coefficients $\alpha(\bar{a}^{s}_{\text{eff}})_{\mu}$ and $\bar{c}^{s}_{00}$ are dependent on the particle decomposition of the source that respectively are given by
\begin{eqnarray}
 \alpha(\bar{a}^{s}_{\text{eff}})_{\mu} &=& \sum_{w} N^{w}_{s}\alpha(\bar{a}^{w}_{\text{eff}})_{\mu} \\
  \bar{c}^{s}_{00} &=& \frac{1}{M}\sum_{w} N^{w}_{s}m^{w}\bar{c}^{w}_{00}
\end{eqnarray}
where the superscript $w=e,p$ or $n$ represents electron, proton, or neutron, $m^{w}$ is the mass of particle $w$, and $N^{w}_{s}$ is the number of particles $w$ of the gravitational source $s$. Considering the Lorentz violations induced by neutral macroscopic bodies, the Lorentz-violating effects are sensitive to the combinations of electron and proton $\alpha(\bar{a}^{e+p}_{\text{eff}})_{\mu}=\alpha(\bar{a}^{e}_{\text{eff}})_{\mu}+\alpha(\bar{a}^{p}_{\text{eff}})_{\mu}$ and $\bar{c}^{(e+p)}_{00}=\bar{c}^{e}_{00}+\bar{c}^{p}_{00}$. Neglecting the higher multipoles, two potentials are given by
\begin{eqnarray}\label{am1}
  U&=&\frac{G_0 M}{r},\nonumber\\
  U^{ij}&=&\frac{G_0 M r^{i}r^{j}}{r^{3}},
\end{eqnarray}
where $G_{0}$ is the Newton gravitational constant. Eq.(\ref{am}) gives the metric for a single body. When considering multiple gravitational sources, the metric is given by the superposition of all gravitational-source potentials and corresponding Lorentz-violating coefficients. In the case of general relativity, the coefficients $\bar s^{\mu\nu}$, $(\bar{a}^{s}_{\text{eff}})_{\mu}$ and $\bar{c}^{s}_{00}$ vanish.

\begin{figure}
\includegraphics[width=0.4\textwidth]{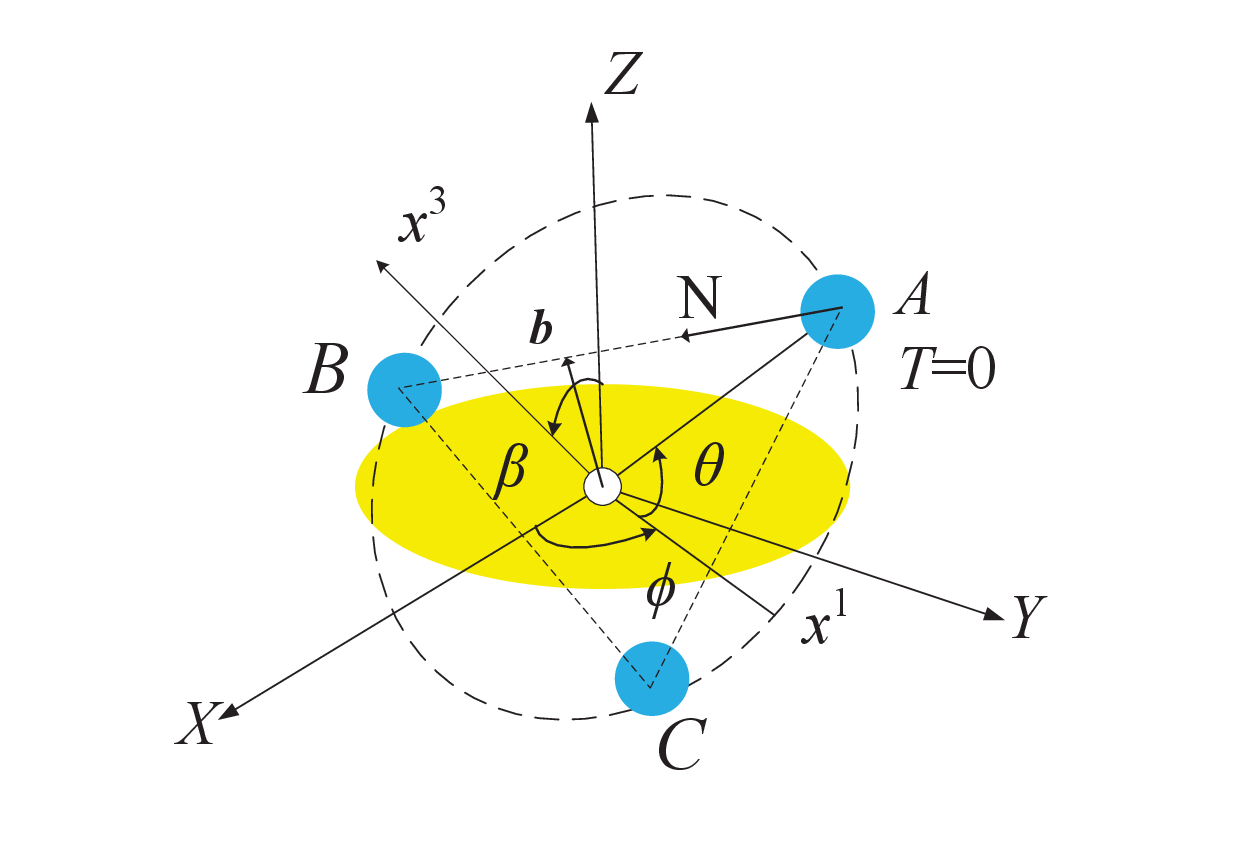}
\caption{ Satellite orbital parameter in the Sun-centered frame. The $( X, Y, Z)$ frame is the Sun-centered inertial reference frame and $( x^1, x^2, x^3)$ frame is the orbital coordinate system. Two frames are associated with two angles, the longitude of the node $\phi$ and the orbital inclination $\beta$. The angles $(\phi,\beta,\theta)$ are the Euler angles for satellite $A$. $\textbf{N}$ is unit vector pointing from $A$ to $B$, and $\textbf{\emph{b}}$ is the corresponding impact parameter vector. }\label{fig.cs}
\end{figure}

For the experimental analyses, the results generally are expressed in the Sun-centered inertial reference frame, $(cT, X, Y, Z)$, with the $ X$ axis pointing from the Sun toward the vernal equinox, the $Z$ axis parallelling the direction of the Earth's rotation, and a right-handed relationship for the $ Y$ axis ($T$ denotes the time coordinate with its origin, $T=0$, at the 2000 vernal equinox.)
For the GW detectors, three satellites constitute an equilateral triangular configuration and it is more convenient to calculate the Lorentz-violating effects in a local reference system. We introduce the orbital coordinate system $( x^1, x^2, x^3)$, where the $x^1$ axis is along the descending node, the $x^3$ axis is along the direction of the orbital angular moment of the satellite, and $ x^2= x^3\times  x^1$. For the geocentric option, the center of the orbit is at the center-of-mass of the Earth. For the heliocentric option, the orbital center is at the center of the satellite's triangular configuration. As shown in FIG. \ref{fig.cs}, in the Sun-centered frame, the orientation of the orbit can be described by two angles, the longitude of the node $\phi$ and the orbital inclination $\beta$. In the orbital frame, the satellite's position is determined by a circular-orbit phase $(\omega t+\theta)$, where $\omega$ is the mean satellite frequency and $\theta$ is the angle between the line of ascending code and the initial position of the satellite.

Neglecting the boost terms, the transformation from the Sun-centered frame to the orbital frame can be expressed as the rotation matrix associated with these two angles
\begin{eqnarray}\label{tansxX}
\left(
  \begin{array}{c}
     x^1 \\
     x^2 \\
     x^3
  \end{array}
  \right)&=&
  {\cal R}\left({\phi,\beta}\right)\left(
  \begin{array}{c}
     X\\
     Y \\
     Z
  \end{array}
  \right) \,,
  \end{eqnarray}
where the rotation matrix is given by
\begin{eqnarray}\label{matrixos}
{\cal R} &=&
  \left(
  \begin{array}{ccc}
    \cos\phi& \sin\phi& 0 \\
    -\sin\phi\cos\beta &\cos\phi\cos\beta & \sin\beta\\
    \sin\phi\sin\beta\ & -\cos\phi\sin\beta & \cos\beta
  \end{array}
  \right) \,.
\end{eqnarray}

\section{The Lorentz-violating effects in the space-based GW missions}\label{sct3}

In the framework of SME, we focus on Lorentz-violating effects of the pure-gravity and matter-gravity coupling sectors. Owing to operating with precision instruments, the space-based GW detectors can be used to probe several classic tests for Lorentz violation effects, including gravitational time delay of light and frequency shift of clock. Considering the linear gravity limits, the leading corrections to classic experiments have been given some discussions in the post-Newtonian coordinate system \cite{PhysRevD.74.045001,PhysRevD.80.044004,PhysRevD.84.085025}. The dominant effects of Lorentz violation in these experiments are sensitive to different combinations of the SME coefficients. In the space-based GW missions, the signals and attainable sensitivities for Lorentz-violating coefficients are presented in the following.

The indices 1 and 2 refer to projections on the orbital plane. The labels 1 and 2 represent the projection onto the line of ascending nodes and the projection onto the perpendicular direction in the orbital plane, respectively. From the transformation matrix, the combinations of SME coefficients can be expressed in terms of the bias Sun-centered coefficients and the explicit expressions of these combinations are
\begin{eqnarray}\label{a4}
{{\bar s}^{11}} - {{\bar s}^{22}}&=&  \left( {{{\cos }^2}\phi  - {{\sin }^2}\phi {{\cos }^2}\beta } \right){{\bar s}^{XX}} \nonumber\\
 &+& \left( {{{\sin }^2}\phi  - {{\cos }^2}\phi {{\cos }^2}\beta } \right){{\bar s}^{YY}} - {\sin ^2}\beta {{\bar s}^{ZZ}}    \nonumber\\
&+&  2\sin \phi \cos \phi \left( {1 + {{\cos }^2}\beta } \right){{\bar s}^{XY}}   \nonumber\\
&+& 2\sin \beta \cos \beta \sin \phi {{\bar s}^{XZ}} - 2\sin \beta \cos\beta \cos\phi {{\bar s}^{YZ}},
\end{eqnarray}

\begin{eqnarray}\label{as412}
{{\bar s}^{11}} + {{\bar s}^{22}}&=& \left( {{{\cos }^2}\alpha  + {{\sin }^2}\alpha {{\cos }^2}\beta } \right){{\bar s}^{XX}} \nonumber\\
 &+& \left( {{{\sin }^2}\alpha  + {{\cos }^2}\alpha {{\cos }^2}\beta } \right){{\bar s}^{YY}} + {\sin ^2}\beta {{\bar s}^{ZZ}}    \nonumber\\
&+&  2\sin \alpha \cos \alpha \left( {1 - {{\cos }^2}\beta } \right){{\bar s}^{XY}}   \nonumber\\
&-& 2\sin \beta \cos \beta \sin \alpha {{\bar s}^{XZ}} + 2\sin \beta \cos\beta \cos\alpha {{\bar s}^{YZ}},
\end{eqnarray}

\begin{eqnarray}\label{a5}
  {\bar s^{12}} &=&  - \sin \phi \cos \phi \cos \beta \left( {{{\bar s}^{XX}} - {{\bar s}^{YY}}} \right) \nonumber\\
 &+&  \left( {{{\cos }^2}\phi  - {{\sin }^2}\phi } \right)\cos \beta {\bar s^{XY}} \nonumber\\
  &+& \cos \phi \sin\beta {\bar s^{XZ}} + \sin \phi \sin \beta {\bar s^{YZ}},
\end{eqnarray}

\begin{equation}\label{a6}
  {\bar s^{01}} = \cos \phi {\bar s^{TX}} + \sin \phi {\bar s^{TY}},
\end{equation}

\begin{equation}\label{a7}
{\bar s^{02}} =  - \sin \phi \cos \beta {\bar s^{TX}} + \cos \phi \cos \beta {\bar s^{TY}} + \sin \beta {\bar s^{TZ}},
\end{equation}

\begin{equation}\label{aa6}
  \alpha{(\bar{a}^{s}_{\text{eff}})^{1}} = \cos \phi \alpha{(\bar{a}^{s}_{\text{eff}})^{X}}  + \sin \phi \alpha{(\bar{a}^{s}_{\text{eff}})^{Y}} ,
\end{equation}

\begin{equation}\label{aa7}
\alpha{(\bar{a}^{s}_{\text{eff}})^{2}} =  - \sin \phi \cos \beta \alpha{(\bar{a}^{s}_{\text{eff}})^{X}}  + \cos \phi \cos \beta \alpha{(\bar{a}^{s}_{\text{eff}})^{Y}}  + \sin \beta \alpha{(\bar{a}^{s}_{\text{eff}})^{Z}} .
\end{equation}

From these equations, the experimental sensitivity for SME coefficients is dependent on the orientation of the orbital plane. The most famous test is from the lunar laser ranging. In the case of lunar laser ranging, the ranging precision of the subcentimeter level has been achieved. By using a set of over 20,000 normal points covering the period of 1969 to 2016, the SME coefficients have been put with stringent constraints \cite{PhysRevLett.117.241301,PhysRevLett.119.201102}. As ranging accuracy improves, better bounds on the coefficients for Lorentz violation can be expected.
For artificial satellites, the orbital orientation is diverse and can be modulated during the operation. The measurements of artificial satellites have the great potential to set the constraints on independent combinations of the coefficients $\bar{s}^{JK}$ that are insensitive in the lunar laser ranging. Considering the polar orbit of the satellite with inclination $\beta=\pi/2$, the $2\omega$ frequency bands are sensitive to all six coefficients $\bar{s}^{JK}$. For example, for $\phi=0$, the satellite laser ranging is sensitive to $\bar{s}^{11}-\bar{s}^{22}=\bar{s}^{XX}-\bar{s}^{ZZ}, \bar{s}^{12}=\bar{s}^{XZ}$, and for $\phi=\pi/2$, the satellite laser ranging is sensitive to $\bar{s}^{11}-\bar{s}^{22}=\bar{s}^{YY}-\bar{s}^{ZZ}, \bar{s}^{12}=\bar{s}^{YZ}$.

\subsection{Time delay of light and Lorentz violation}\label{sec1time}
In space-based GW detectors, laser interferometer can provide accurate information of light travel time including gravitational time delay. Lorentz-violating corrections to the Shapiro time delay have the potential to yield a constraint for SME coefficients. Once the spacetime metric is given, the standard method for studying the light propagation in a gravitational field is to solve the null geodesic equation or the eikonal equation \cite{ashby2010accurate,weinberg1973gravitation,serbenta2022bilocal}. For the light propagation in the Solar-System case, many solutions have been proposed in the post-Newtonian and post-Minkowskian approximations \cite{klioner2003practical,crosta2006microarcsecond,zschocke2010efficient,PhysRevD.90.084020,PhysRevD.100.064063}. Moreover, a different method is available, which is based on the time transfer function \cite{le2004world,teyssandier2008general,PhysRevD.89.064045,teyssandier2022time,PhysRevD.93.044028}. This formalism can be used to compute the light travel coordinate time between the emission spacetime point and the reception spacetime point. In the computation of the gravitational time delay, we in addition consider the contributions from the pure gravitational and matter-gravity coupling sectors of the minimal SME. We consider that a one-way laser signal is sent from a point event $A$ $(ct_{A},\textbf{\emph{x}}_{A})$ to another point event $B$ $(ct_{B},\textbf{\emph{x}}_{B})$. The time transfer function between point events is expressed as $T_{AB}=R_{AB}/c+\Delta_{g}(A,B) /c$, where the $\Delta_{g}(A,B)$ is the gravitational delay function. The time transfers function satisfies the Hamilton-Jacobi-like function, then one can deduce the gravitational delay function from the metric (\ref{am}). The metric can be decomposed as $g_{\mu\nu}=\eta_{\mu\nu}+h_{\mu\nu}$ with $\eta_{\mu\nu}=$diag($-$1,+1,+1,+1) and $h_{\mu\nu}$ corresponds to the $c^{-2}$ terms of metric (\ref{am}). To the post-Newtonian order, the delay function $\Delta_{g}(A,B)$ is given by the integro-differential equation \cite{teyssandier2008general}
\begin{equation}\label{gtime0}
  \Delta_{g}(A,B)=\frac{R_{AB}}{2}\int_{0}^{1}\left( h_{00}+2h_{0i}N^{i}_{AB}+h_{ij}N^{i}_{AB}N^{j}_{AB} \right)d\lambda,
\end{equation}
where $R_{AB}=|\textbf{\emph{x}}_{B}-\textbf{\emph{x}}_{A}|$ is the coordinate distance from emission point to reception point, $N^{i}_{AB}=(\textbf{\emph{x}}_{B}^{i}-\textbf{\emph{x}}_{A}^{i})/R_{AB}$, and the integral is calculated along the straight line between $A$ and $B$ that is defined by the parametric equation
\begin{equation}\label{dxlam}
  x^{i}=(x^{i}_{B}-x^{i}_{A})\lambda+x^{i}_{A},\,\,\,\,0\leq\lambda\leq1,
\end{equation}
where $\lambda$ is an affine parameter. We use a rescaled observable Newtonian constant defined as $G=G_{0}(1+5{\bar s}^{00}/3)$.  After some calculations, the expression for the one-way light travel time is given by \cite{PhysRevD.80.044004,PhysRevLett.119.201102}
\begin{eqnarray}\label{gtime1}
  T_{AB}
&=& \frac{R_{AB}}{c} + \frac{{2GM}}{{{c^3}}}\left( {1 - \frac{2}{3}{{\bar s}^{00}}+\bar{c}^{s}_{00}+\frac{\alpha}{M}(\bar{a}^{s}_{\text{eff}})^{0} -\left[ {{\bar s}^{0j}}-\frac{\alpha}{M}(\bar{a}^{s}_{\text{eff}})^{j} \right] {{N}^j}} \right)\ln \frac{{{r_A} + {r_B} + R_{AB}}}{{{r_A} + {r_B} - R_{AB}}} \nonumber\\
 &+& \frac{{GM}}{{{c^3}}} \big{[} - {{\bar s}^{00}}- \frac{\alpha}{M}(\bar{a}^{s}_{\text{eff}})^{0} + \left[{{\bar s}^{0j}}-\frac{\alpha}{M}(\bar{a}^{s}_{\text{eff}})^{j} \right]{{N}^j} + {{\bar s}^{jk}}{{\hat b}^j}{{\hat b}^k}
 \big{]} \left( \textbf{\emph{n}}_{B}\cdot\textbf{N}-\textbf{\emph{n}}_{A}\cdot\textbf{N} \right) \nonumber\\
&+&  \frac{{GM}}{{{c^3}}}\left( {\left[{{\bar s}^{0j}}-\frac{\alpha}{M}(\bar{a}^{s}_{\text{eff}})^{j}\right] {\hat b^j} -{{\bar s}^{jk}}{{N}^j}{\hat b^k}} \right)\frac{b({{r_A} - {r_B}})}{{{r_A}{r_B}}},
\end{eqnarray}
where the unit vectors are $\textbf{\emph{n}}_{A}=\textbf{\emph{x}}_{A}/|\textbf{\emph{x}}_{A}|$, $\textbf{\emph{n}}_{B}=\textbf{\emph{x}}_{B}/|\textbf{\emph{x}}_{B}|$ and $\textbf{N}=(\textbf{\emph{x}}_{B}-\textbf{\emph{x}}_{A})/|\textbf{\emph{x}}_{B}-\textbf{\emph{x}}_{A}|$, ${N}^i$ is the $i$-th component of unit vector $\textbf{N}$ (as shown in Fig.\ref{fig.cs}), $b^i$ is the impact parameter vector with ${\hat b}^i=b^i/b$ $(\textbf{\emph{b}} = \textbf{N} \times (\textbf{\emph{x}}_{A} \times \textbf{N}))$, and $r_{A/B}=|\textbf{\emph{x}}_{A/B}|$ is the distance between point $A/B$ to center-of-mass of gravitational body.

In space-based missions, three satellites (satellites $A$, $B$, and $C$) form an equilateral triangle configuration, which leads to three laser arms and six GW data streams. Six data streams can be used to construct various measurement combinations.
Considering the measurement of laser interferometry, the laser signal received on the satellite $B$ is coherently retransmitted back to satellite $A$. The measurement signal involves the round-trip light travel time that can be established by two data streams (one-way signal and return-trip signal). The return-trip time of the laser signal can be obtained from Eq.(\ref{gtime1}) by exchanging the $A$ and $B$ labels. Since the satellite's velocity is much smaller than the speed of light, the $GMv/c^{4}$ terms in the return-trip time can be neglected.
Thus, gravitational time delay in the round-trip light travel time is given by
\begin{eqnarray}\label{gtime2}
   \Delta \tau_{g}(A,B)
&=&  \frac{{4GM}}{{{c^3}}} \left(  1 - \frac{2}{3} {{\bar s}^{00}}+\bar{c}^{s}_{00}+\frac{\alpha}{M}(\bar{a}^{s}_{\text{eff}})^{0} \right) \ln \frac{{{r_A} + {r_B} + R_{AB}}}{{{r_A} + {r_B} - R_{AB}}}\nonumber\\
 &+&\frac{{2GM}}{{{c^3}}} \big{[} - {{\bar s}^{00}}-\frac{\alpha}{M}(\bar{a}^{s}_{\text{eff}})^{0}  + {{\bar s}^{jk}}{{\hat b}^j}{{\hat b}^k} \big{]} \left( \textbf{\emph{n}}_{B}\cdot\textbf{N}-\textbf{\emph{n}}_{A}\cdot\textbf{N} \right)
-  \frac{{2GM}}{{{c^3}}} {{\bar s}^{jk}}{{N}^j}{\hat b^k}\frac{b({{r_A} - {r_B}})}{{{r_A}{r_B}}}.
\end{eqnarray}
This kind of round-trip path is called a single-arm round-trip path. It should be noted that when considering the contributions of outgoing and return signals, the terms involving the ${{\bar s}^{0i}}$ and $\alpha(\bar{a}^{s}_{\text{eff}})^{i}$ coefficients canceled out since these coefficients are parity-odd coefficients. As a result, the single-arm round-trip signals make the experiment sensitive to different combinations of SME coefficients. For the convenience of plotting in the following section, we mark the first and second terms of Eq.(\ref{gtime2}) as Lv1 and Lv2, respectively.

$\emph{The Earth's gravitational effect:}$ Firstly, we consider the influences of Earth's gravitational time delay on the space-based GW detectors. Due to the large distance between the GW detectors and the Earth in the heliocentric option, such as LISA and Taiji, the influence of the Earth's gravity is minimal. Therefore, we mainly focus on the influences of the Earth's gravitational time delays on the geocentric GW detectors, using the TianQin mission as an example. In the TianQin mission, three satellites form an equilateral triangle configuration, with three laser arms connecting them. Each arm is almost the same length, approximately $1.73\times10^{8}$ m, and the angle between the two arms is $60^{\circ}$.
The gravitational delay of single-arm round-trip light signal in the other two arms can be obtained by using Eq.(\ref{gtime2}) with corresponding substitutions for the symbols $A$ and $B$. The gravitational-wave interferometers provide not only inter-satellite ranging but also interference measurements. The interference signal in the gravitational-wave detector is derived from the phase difference of the laser light signal in the adjacent two arms. The measurement signal in the satellite $A$ can be obtained by using the round-trip light phases from satellites $B$ and $C$, $\delta\varphi=\Delta\varphi(t,A,B)-\Delta\varphi(t,A,C)$, where $\Delta\varphi(t,A,B/C)$ is the round-trip light phases between satellites $A$ and $B/C$. Due to the symmetrical configuration relative to Earth, we have the following relationships, $(\textbf{\emph{n}}_{B}-\textbf{\emph{n}}_{A})\cdot\textbf{N}(A,B)\cong(\textbf{\emph{n}}_{C}-\textbf{\emph{n}}_{A})\cdot\textbf{N}(A,C)$, $r_{A}\cong r_{B}\cong r_{C}$, and the angle between the $\textbf{\emph{n}}_{A}$ and $\textbf{\emph{b}}$ is $\pi/3$ in the orbital frame. Then, the Lorentz-violating term in the interference signal is
\begin{eqnarray}\label{df}
\delta \tau_{g,E}&=&  \Delta \tau_{g}(A,B)-\Delta\tau_{g}(A,C)  \nonumber\\
&=& \frac{3GM_{E}}{c^{3}} \big{[}\left( \bar s^{22}-\bar s^{11}\right)\sin\left(2\omega t+2\theta+4\pi/3\right)
  +2\bar s^{12} \cos\left(2\omega t+2\theta+4\pi/{3}\right)\big{]},
\end{eqnarray}
where $M_{E}$ is the Earth's mass, the logarithmic term is canceled due to the symmetrical light paths with respect to Earth. The interference signals are sensitive to the combinations of SME coefficients $\bar s^{22}-\bar s^{11}$ and $\bar{s}^{12}$. In this equation, the terms dominated by coefficients $\bar s^{22}-\bar s^{11}$ and $\bar{s}^{12}$ are labeled Lv3 and Lv4, respectively.

In addition, the three single-arm round-trip light signals can form a triangular closed-loop light path, which is a unique measurement in space-based GW detectors. To distinguish it from the single-arm case, we call this path a triangular round-trip path.
In the orbital frame, the gravitational delay signal in the triangular round-trip path is expressed as
\begin{eqnarray}\label{closetwo}
  \Delta \tau_{g,E}^{\text{cl}}&=&\Delta \tau_{g}(A,B)+\Delta\tau_{g}(B,C)+\Delta\tau_{g}(C,A)\nonumber\\
  &\cong&\frac{{12GM_{E}}}{{{c^3}}} \left( 1 -\frac{2}{3} {{\bar s}^{00}}  +\bar{c}^{E}_{00}+\frac{\alpha}{M_E}(\bar{a}^{E}_{\text{eff}})^{0} \right) \ln \frac{{{r_A} + {r_B} + R_{AB}}}{{{r_A} + {r_B} - R_{AB}}}\nonumber\\
  &+&\frac{3GM_E}{c^{3}}\left[  - 2{{\bar s}^{00}}-2\frac{\alpha}{M_E}(\bar{a}^{E}_{\text{eff}})^{0}+(\bar s^{11}+\bar s^{22})\right] \left(\textbf{\emph{n}}_{B}\cdot\textbf{N}-\textbf{\emph{n}}_{A}\cdot\textbf{N} \right),
\end{eqnarray}
where the composition of the Earth is characterized by $N^{e}_{E}/M_{E}=N^{p}_{E}/M_{E} \approx N^{n}_{E}/M_{E} \approx 0.5$ (GeV/$c^{2}$)$^{-1}$ and $\bar{c}^{E}_{00}\approx0.5\bar{c}^{(e+p)}_{00}+0.5\bar{c}^{n}_{00}$. The coefficients $\bar s^{ij}$ with $(i\neq j)$ cancelled when considering the closed-loop light path, which is the unique feature of the equilateral triangle configuration. The closed-loop light measurement can provide a new combination of SME coefficients $\bar s^{11}+\bar s^{22}-2\bar{s}^{00}-2{\alpha}(\bar{a}^{E}_{\text{eff}})^{0}/{M_E}$ (for the convenience of plotting in following section, we label this term Lv5), which leads to a new coefficients combination $\bar s^{11}+\bar s^{22}-2\bar{s}^{00}$ for pure gravity sector. By using the linear combination $\bar s^{11}-\bar s^{22}$ in the interference signal (\ref{df}), one can derive independent constraints on the individual Lorentz violation coefficients $\bar s^{11}$ and $\bar s^{22}$.

Apart from the combinations of the round-trip light signals, space-based GW missions can also utilize six data streams between three satellites to construct combinations of one-way light signals. These signal combinations provide sensitivity to the different combinations of SME coefficients from those probed by the round-trip signals. Considering one-way signals $AB$ and $BA$, it yields a measurement for the temporal-spatial component ${{\bar s}^{0j}}$ of pure-gravity coefficients and spatial component $\alpha(\bar{a}^{w}_{\text{eff}})^{j}$ of matter-gravity coupling coefficients
\begin{eqnarray}\label{gtime122}
\Delta \tau_{g,E}(A,B)-\Delta\tau_{g,E}(B,A)&=& \frac{{2GM_E}}{{{c^3}}}\left[ { \left(\frac{\alpha}{M_E}(\bar{a}^{s}_{\text{eff}})^{j}- {{\bar s}^{0j}} \right) {{N}^j}} \right]\ln \frac{{{r_A} + {r_B} + R_{AB}}}{{{r_A} + {r_B} - R_{AB}}} \nonumber\\
 &+& \frac{{GM_E}}{{{c^3}}} \left[   \left({{\bar s}^{0j}}-\frac{\alpha}{M_E}(\bar{a}^{s}_{\text{eff}})^{j} \right){{N}^j}
  \left( \textbf{\emph{n}}_{B}\cdot\textbf{N}-\textbf{\emph{n}}_{A}\cdot\textbf{N} \right)
 + {\left({{\bar s}^{0j}}-\frac{\alpha}{M_E}(\bar{a}^{s}_{\text{eff}})^{j}\right) {\hat b^j} } \frac{b({{r_A} - {r_B}})}{{{r_A}{r_B}}}\right].\nonumber\\
\end{eqnarray}
Overall, using the combinations of six data streams, it is possible to construct various combinations for limiting Lorentz-violating coefficients.

$\emph{The Sun's gravitational effect:}$ For the Sun's gravitational contribution, the time delays in the one-way and single-arm round-trip light paths can be respectively obtained by reexpressing Eqs.(\ref{gtime1}) and (\ref{gtime2}) with all quantities replaced by the Sun's quantities. However, the GW detector is far from the Sun and the satellite-Sun distance is about 1 AU, far greater than the distances between satellites or laser arms. The influences of the Sun's gravitational delays on the heliocentric and geocentric options of GW detectors are at a similar level. Therefore, we will consider the Sun's influences on the classic space-based GW detectors, such as LISA, Taiji, and TianQin. These GW missions satisfy the condition $r_{A}, r_{B} \gg R_{AB}$, and the logarithmic term of the Sun's gravitational delay can be approximated as $2R_{AB}/(r_{A}+r_{B})$.
The Sun's gravitational delay time in the single-arm round-trip signal can be expressed as
\begin{eqnarray}\label{gtimes2}
   \Delta \tau_{g,S}(A,B) &\cong&  \frac{{8GM_{S}}}{{{c^3}}} \left(  1 - \frac{2}{3} {{\bar s}^{00}}+\bar{c}^{S}_{00}+\frac{\alpha}{M_S}(\bar{a}^{S}_{\text{eff}})^{0}  \right)  \frac{{R_{AB}}}{{{r_A} + {r_B} }} \nonumber\\
 &+& \frac{{2GM_{S}}}{{{c^3}}} \left[ - {{\bar s}^{00}}-\frac{\alpha}{M_S}(\bar{a}^{S}_{\text{eff}})^{0} + {{\bar s}^{jk}}{{\hat b}^j}{{\hat b}^k} \right] \left(\textbf{\emph{n}}_{B}-\textbf{\emph{n}}_{A} \right) \cdot\textbf{N} -\frac{{2GM_{S}}}{{{c^3}}} {{\bar s}^{jk}}{{N}^j}{\hat b^k}\frac{b({{r_A} - {r_B}})}{{{r_A}{r_B}}},
\end{eqnarray}
where $M_S$ represents the Sun's mass, the composition of the Sun is considered as $N^{e}_{S}/M_{S}=N^{p}_{S}/M_{S} \approx 0.9$ (GeV/$c^{2}$)$^{-1}$, $N^{n}_{S}/M_{S} \approx 0.1$ (GeV/$c^{2}$)$^{-1}$ and $\bar{c}^{S}_{00}\approx0.9\bar{c}^{(e+p)}_{00}+0.1\bar{c}^{n}_{00}$, and all quantities are expressed with respect to the Sun in the heliocentric-ecliptic coordinate system. The Sun's gravitational delays are sensitive to the matter-gravity coupling coefficients $\alpha(\bar{a}^{(e+p)}_{\text{eff}})^{0}$ and $\bar{c}^{(e+p)}_{00}$ of SME.

For the effects of Sun gravity, the interference signals of relativistic gravitational delays at spacecraft $A$ are given by $\Delta \tau_{g}(A,B)-\Delta \tau_{g}(A,C)$, which can be split into the time-component part and spatial-component part
\begin{equation}\label{spts}
  \delta \tau_{g,S} =\delta \tau^{t}_{g,S} +\delta \tau^{s}_{g,S}
\end{equation}
with
\begin{eqnarray}\label{gtimesin1}
  \delta \tau^{t}_{g,S} &=&  \left( 1 - \frac{2}{3} {{\bar s}^{00}}+\bar{c}^{S}_{00}+\frac{\alpha}{M_S}(\bar{a}^{S}_{\text{eff}})^{0}  \right)\frac{{8GM_{S}}}{{{c^3}}} \left(\frac{R_{AB}}{r_{A}+r_{B}}-\frac{R_{AC}}{r_{A}+r_{C}}\right)\nonumber\\
   &-&\left( \bar{s}^{00} +\frac{\alpha}{M_S}(\bar{a}^{S}_{\text{eff}})^{0}\right) \frac{{2GM_S}}{{{c^3}}}\left[ \left(\textbf{\emph{n}}_{B}-\textbf{\emph{n}}_{A} \right) \cdot\textbf{N}-\left(\textbf{\emph{n}}_{C}-\textbf{\emph{n}}_{A} \right) \cdot\textbf{N}'\right],
\end{eqnarray}
and
\begin{eqnarray}\label{sss1}
  \delta \tau^{s}_{g,S} &=&\frac{{2GM_{S}}}{{{c^3}}}\left[\left( {{\bar s}^{jk}}{{\hat b}^j}{{\hat b}^k}\right)_{AB}\left(\textbf{\emph{n}}_{B}-\textbf{\emph{n}}_{A} \right) \cdot\textbf{N}- \left({{\bar s}^{jk}}{{\hat b}^j}{{\hat b}^k}\right)_{AC}\left(\textbf{\emph{n}}_{C}-\textbf{\emph{n}}_{A} \right) \cdot\textbf{N}'\right] \nonumber\\
  &-& \frac{{2GM_{S}}}{{{c^3}}} \left[ \left({{\bar s}^{jk}}{{N}^j}{ b^k}\right)_{AB} \frac{{{r_A} - {r_B}}}{{{r_A}{r_B}}}- \left({{\bar s}^{jk}}{{N'}^j}{ b^k}\right)_{AC}\frac{{{r_A} - {r_C}}}{{{r_A}{r_C}}}\right],
\end{eqnarray}
where $\textbf{N}'=(\textbf{\emph{x}}_{C}-\textbf{\emph{x}}_{A})/|\textbf{\emph{x}}_{C}-\textbf{\emph{x}}_{A}|$, and subscript $AB(AC)$ represents the signal link $AB(AC)$. For the time-component part (\ref{gtimesin1}), the first term contains the contributions from the leading-order GR Shapiro delay (or logarithmic term) and time-component coefficients modification of Lorentz violation. This term is much small since the nearly equal lengths of two interference arms result in that gravitational delays are almost canceled out in the interference signal. The second term in Eq.(\ref{gtimesin1}) is contributed from the nonlogarithmic term of time-component coefficients. This term is an enhanced term in interference signal and is much greater than the influence of the first term (logarithmic term). The time-component coefficients $\bar{s}^{00}$, $\alpha(\bar{a}^{w}_{\text{eff}})^{0}$ of Lorentz violation are sensitive to the nonlogarithmic terms in interference signals. Eq.(\ref{sss1}) represents the Lorentz-violating signals controlled by the spatial-component coefficients $\bar{s}^{ij}$.

Considering the triangular round-trip path, the gravitational-time-delay signal due to the Sun's gravity can be expressed similarly as $\Delta \tau_{g}(A,B)+\Delta\tau_{g}(B,C)+\Delta\tau_{g}(C,A)$, where all terms are expressed with respect to the Sun. This closed-loop combination can be used to analyze the influences of the Sun's gravitational delays and corresponding Lorentz-violating delays to space-based GW detectors. And the combinations of one-way light signals can be used to limit the temporal-spatial component ${{\bar s}^{0j}}$ of pure-gravity coefficients and spatial component $\alpha(\bar{a}^{w}_{\text{eff}})^{j}$ of matter-gravity coupling coefficients. Furthermore, there are additional features in the geocentric GW missions. For the TianQin, Earth is inside the closed-loop light path formed by the three satellites, and the Sun is outside this closed-loop path. Considering the Sun's influences on the round-trip light of a single arm, the time-delay signals present a change with the frequency $\omega$, which is caused by the satellite's orbit around the Earth. For the closed-loop path, the time-delay signals of the Sun's gravity do not exhibit a change in frequency of $\omega$ (as shown in FIG. \ref{fig41}). For GR effects or Lorentz-violating effects in the closed-loop light path, the orbit-frequency variations would be averaged out, and the time delays in the closed-loop path are insensitive to the orbital frequency $\omega$.

For space-based GW missions, the laser interferometers provide extremely hight sensitivity to displacement measurement. Considering classic geocentric and heliocentric GW detectors, the displacement measurement accuracy is given by an empirical criterion \cite{luo2016tianqin,jennrich2009lisa,armano2009lisa,danzmann2003lisa,PhysRevLett.120.061101,luo2021taiji,luo2020brief}
\begin{equation}\label{htdl1}
  S^{1/2}_{x} \approx \frac{1\text{pm}}{\sqrt{2}\sqrt{\text{Hz}}}\sqrt{1+\left(\frac{f_{i}}{f}\right)^{4}},
\end{equation}
where $f$ is the frequency of measurement signal, $f_{i}$ is given by $10$ mHz in the TianQin and $3$ mHz in the LISA and Taiji. TianQin detector reaches a position sensitivity 1 pm/Hz$^{1/2}$ at $\sim$10 mHz, and the sensitivity of LISA and Taiji detector can reach of 1 pm/Hz$^{1/2}$ at $\sim$3 mHz. For the frequencies where the SME coefficients are sensitive, the precision of displacement measurement can reach $\sim 10^{-6}$ m/Hz$^{1/2}$. This implies that gravitational-delay measurements have potential to set limit on SME coefficients $\bar{s}_{TT}$, $\alpha (\bar{a}^{(e+p)}_{\text{eff}})_{T}$, and $\bar{c}^{(e+p)}_{TT}$ with the level of $10^{-6}$, and on SME coefficients $\alpha (\bar{a}^{n}_{\text{eff}})_{T}$, and $\bar{c}^{n}_{TT}$ with the level of $10^{-5}$.
Although, the frequencies of the Lorentz-violating signals are several orders of magnitude below the optimized band of the instruments, the option for sidestepping this issue is to use the information circulating at sideband frequencies, which can be used to measure the low-frequency Lorentz-violating signals \cite{kostelecky2016searching,gusev2014low}.

\begin{table}[!t]
\caption{\label{tab:2} The estimated sensitivity of SME coefficients. The index $T$ indicates that the limits are held in the Sun-centered inertial reference frame.}
\newcommand{\tabincell}[2]{\begin{tabular}{@{}#1@{}}#2\end{tabular}}
\begin{tabular}{cccccccc}
\hline
\hline
\tabincell{l}
SME Coefficients\,\,\,\,\,\,\,  &$\alpha (\bar{a}^{n}_{\text{eff}})_{T}$ \,\,\,\,\,\,\,  & $\alpha (\bar{a}^{(e+p)}_{\text{eff}})_{T}$ \,\,\,\,\,\,\,  &$\bar{c}^{n}_{TT}$\,\,\,\,\,\,\,      &$\bar{c}^{(e+p)}_{TT}$   \,\,\,\,\,\,\,  &$\bar{s}_{TT}$  \,\,\,\,\,\,\,   &$\bar{s}_{TJ}$ \,\,\,\,\,\,\,  &$\bar{s}_{JK}$\\
 &(GeV)\,\,\,\,\,\,\, &(GeV)\,\,\,\,\,\,\, &\,\,\,\,\,\,\, & \,\,\,\,\,\,\,  &  \,\,\,\,\,\,\,  &   \,\,\,\,\,\,\, & \\
\hline
Limit
  &$10^{-5}$ &$10^{-6}$    &$10^{-5}$    &$10^{-6}$  &$10^{-6}$ &$10^{-6}$  &$10^{-6}$\\
\hline
\hline
\end{tabular}
\end{table}

\subsection{Frequency shift and Lorentz violation}
In addition to the time delay and light bending, the Lorentz-violating terms can also lead to the perturbation in frequency shift. The three satellites of GW detectors are equipped with identical clocks that can provide the measurements of frequency shift. The frequency shift between these clocks contains possible Lorentz-violating effects. We consider the frequency shift of the clocks between satellites $A$ and $B$. At time $T_{A}$, the clock in satellite $A$ sends a light signal with proper frequency $f_{A}$, and at time $T_{B}$, this signal is received by the clock in satellite $B$ with proper frequency $f_{B}$. The one-way frequency shift between two clocks is given by \cite{blanchet2001relativistic,jia2023investigation,qin2023tidal,qin2019relativistic}
\begin{equation}\label{fres1}
  \frac{f_{B}}{f_{A}}=\left( \frac{d \tau_{A}}{d T_{A}}\right)_{T_{A}}\left( \frac{d T_{B}}{d \tau_{B}}\right)_{T_{B}}\left(\frac{d T_{A}}{d T_{B}}\right),
\end{equation}
where $\tau_{A}$ and $\tau_{B}$ are the proper times of clocks in satellites $A$ and $B$, respectively.
The derivative term $d T/d\tau$ is dependent on the state of clock, which can be calculated by the invariance of the Rimannian space-time interval $ds^{2}=g_{\mu\nu}dx^{\mu}dx^{\nu}$. The $dT_{A}/dT_{B}$ term can be determined by using the time transfer function $T_{AB}$. With some manipulation, we obtain the frequency shift up to order $c^{-3}$
\begin{equation}\label{frs2}
  \frac{f_{B}}{f_{A}}=1+\left(\frac{f_{B}}{f_{A}}\right)_{G}+\left(\frac{f_{B}}{f_{A}}\right)_{D}.
\end{equation}
The term labeled $G$ represents the effects arising from gravitational field, which contains the gravitational redshift and Lorentz-violating effects. This term is given by
\begin{eqnarray}\label{fres3}
  \left(\frac{f_{B}}{f_{A}}\right)_{G}&=&\left(1-\frac{1}{6}\bar s^{00}+\bar{c}^{S}_{00}+2\frac{\alpha}{M}(\bar{a}^{S}_{\text{eff}})^{0}  \right)\left(\frac{GM}{c^2 r_{B}}-\frac{GM}{c^2 r_{A}}\right)
  +\frac{1}{2}\bar s^{ij}\left( \frac{GMr^{i}_{B}r^{j}_{B}}{c^{2}r^{3}_{B}}-\frac{GMr^{i}_{A}r^{j}_{A}}{c^{2}r^{3}_{A}}\right)\nonumber\\
&+&\left(\frac{\alpha}{M}(\bar{a}^{s}_{\text{eff}})^{j}-\bar{s}^{0i} \right)\left(\frac{GMv^{i}_{B}}{c^{3}r_{B}}- \frac{GMv^{i}_{A}}{c^{3}r_{A}}\right)
+\left(\frac{\alpha}{M}(\bar{a}^{s}_{\text{eff}})^{k}-\bar{s}^{0k} \right) \left(\frac{GMr_{B}^{i}r_{B}^{k}v_{B}^{i}}{c^{3}r_{B}^{3}} -\frac{GMr_{A}^{i}r_{A}^{k}v_{A}^{i}}{c^{3}r_{A}^{3}}\right).
\end{eqnarray}
The first term in Eq.(\ref{fres3}) includes the time-component coefficients corrections to the standard gravitational redshift. The second term is the contribution of the even-parity coefficients for Lorentz violation. The third and fourth terms represent the influences from the parity-odd coefficients.
%As an estimate of attainable sensitivities for space-based GW detector, we assume that the accuracy of satellite-based clock is $10^{-15}$ and the orbital eccentricity is $0.01$. The gravitational redshift due to Sun mass is the order of $1\times10^{-11}$, and the gravitational redshift due to Earth mass is $8\times10^{-13}$. $A$ $priori$ estimation implies clock's sensitives: parts in $10^{5}$ on the coefficients $\bar{s}^{00}$ and $\bar{s}^{ij}$. Gravitational redshift measurements are insensitive to the SME coefficients $\bar{s}^{0i}$.

The term labeled $D$ represents the Doppler effects, which contains the conventional Doppler shifts and gravitational frequency shift due to the gravitational time delay corrections of light (or the wave vector corrections). By an iterative process with time transfer function, this term is given by
\begin{eqnarray}\label{fres4}
  \left(\frac{f_{B}}{f_{A}}\right)_{D}&=&-\frac{\textbf{N}\cdot \textbf{\emph{v}}_{AB} }{c}
  +\frac{1}{c^{2}}\left[ \frac{\textbf{\emph{v}}_{B}^{2}-\textbf{\emph{v}}^{2}_{A}}{2} -\left(\textbf{N}\cdot \textbf{\emph{v}}_{AB}\right)\left(\textbf{N}\cdot \textbf{\emph{v}}_{A}\right)\right]\nonumber\\
  &-&\frac{1}{c^{3}}\left[ \left(\textbf{N}\cdot \textbf{\emph{v}}_{AB}\right)\left(\frac{\textbf{\emph{v}}_{B}^{2}}{2}-\frac{\textbf{\emph{v}}^{2}_{A}}{2} + \left(\textbf{N}\cdot \textbf{\emph{v}}_{A}\right)^{2}\right) \right]
 -\frac{d\Delta_{g}(A,B)}{cdT_{B}},
\end{eqnarray}
where the first three terms are the Doppler shift due to the motion of satellites, the last term is the gravitational frequency shift with the order of $c^{-3}$, and $\Delta_{g}(A,B)$ is the gravitational time delay. The gravitational frequency shift term can be calculated from the delay function (\ref{gtime1}), and the result is significantly more cumbersome than Eq.(\ref{gtime1}). Since the variations of time delays are much small, a rough estimation suggests that the influence of this term is negligible for Lorentz-violating research. For an estimate of magnitude, we keep only the Shapiro-delay term (logarithmic term) in calculating the gravitational frequency shift
\begin{eqnarray}\label{freg}
  \left(\frac{\delta f}{f}\right)_{\Delta}=\frac{2GM}{c^{3}r_{A}r_{B}}\left( {1 - \frac{2}{3}{{\bar s}^{00}}+\bar{c}^{s}_{00}+\frac{\alpha}{M}(\bar{a}^{s}_{\text{eff}})^{0} } \right)\left[ \frac{(r_{A}+r_{B})\textbf{N}_{AB}\cdot\textbf{\emph{v}}_{AB}}{1+\textbf{\emph{n}}_{A}\cdot\textbf{\emph{n}}_{B}}
  -\frac{(\textbf{\emph{n}}_{A}\cdot\textbf{\emph{v}}_{A}+\textbf{\emph{n}}_{B}\cdot\textbf{\emph{v}}_{B})R_{AB}}{1+\textbf{\emph{n}}_{A}\cdot\textbf{\emph{n}}_{B}}\right].
\end{eqnarray}
For TianQin mission, the Earth-gravity and Sun-gravity contributions do not exceed $1\times10^{-15}$. For LISA or Taiji mission, the contribution of this term is much smaller than $1\times10^{-14}$. The contribution of Eq.(\ref{freg}) is much small for GW missions. However, in the case of solar conjunction (SC) experiments, the contribution of this term becomes significant
\begin{equation}\label{sce}
  \left(\frac{\delta f}{f}\right)^{\text{SC}}_{\Delta}=\left( {1 - \frac{2}{3}{{\bar s}^{TT}}+\bar{c}^{S}_{TT}+\frac{\alpha}{M_S}(\bar{a}^{S}_{\text{eff}})^{T} } \right)\frac{4GM_S}{c^{3}b}\dot{b},
\end{equation}
where $\dot{b}$ is the time derivative of impact parameter $b$. By using this effects, the Cassini solar conjunction experiment obtained the measurement for parameterized post-Newtonian (PPN) parameter $\gamma-1=(2.1\pm2.5)\times10^{-5}$ \cite{bertotti2003test}. This may be equivalent to a constraint accuracy $10^{-5}$ for the SME coefficients $\bar{s}^{TT}$, $\alpha (\bar{a}^{(e+p)}_{\text{eff}})_{T}$, and $\bar{c}^{(e+p)}_{TT}$, and $10^{-4}$ for $\alpha (\bar{a}^{n}_{\text{eff}})_{T}$, and $\bar{c}^{n}_{TT}$. With the development of high-precision time and frequency transfer technologies \cite{PhysRevX.6.021016,PhysRevLett.120.050801,shen2022free}, clock comparisons in future space missions have great potential to improve the constraints on these SME parameters.

\section{Numerical study for Lorentz-violating effects}\label{sct4}

To evaluate the orders of magnitude for the various terms of the gravitational time delay, we make a preliminary model to calculate the satellites orbits of space-based GW missions, TianQin, LISA, and Taiji.
Firstly, we consider a preliminary orbital model for TianQin satellites.
We assume that the satellite orbit around the Earth is the Keplerian orbit with the semimajor axis $a=1\times10^{8}$ m and a small eccentricity $e$ (here it is set $\sim0.01$). At a given time $t_{A}$, the position of satellite $A$ can be described by the angle $\omega_{s}t_{A}+\theta$ in the orbital coordinate system. The three satellites and corresponding orbits are equivalent with the same orbital eccentricity $e$ and inclination $i$, but the orbital semimajor axes of satellites $B$ and $C$ are shifted by the angles $2\pi/3$ and $4\pi/3$, respectively, relative to satellite $A$'s. The satellite mutual distances are approximately constant with a small variation caused by the eccentricity. For estimating the gravitational time delay, we calculate the satellite positions to the second terms in the eccentricity. In the orbital coordinate system, the coordinates of the satellite $A$ are given by \cite{hu2018fundamentals}
\begin{eqnarray}\label{coora}
  x_{A}(t)&=&a\cos(\omega_{s}t+\theta)+\frac{ae}{2}\left( \cos 2(\omega_{s}t+\theta)-3\right)
  -\frac{3ae^{2}}{2}\cos(\omega_{s}t+\theta)\sin^{2}(\omega_{s}t+\theta), \nonumber\\
  y_{A}(t)&=&a\sin(\omega_{s}t+\theta)+\frac{ae}{2} \sin 2(\omega_{s}t+\theta)
  +\frac{ae^{2}}{4}\sin(\omega_{s}t+\theta)(3\cos 2(\omega_{s}t+\theta)-1),\nonumber\\
  z_{A}(t)&=&0,
\end{eqnarray}
where $\omega_{s}$ is the frequency of satellite orbit. Since the orbits of three satellites are require to be equivalent, we can obtain the coordinates of the satellites $B$ and $C$ by corresponding phase shift $2\pi/3$ and $4\pi/3$ in term $(\omega_{s}t+\theta)$ of Eq.(\ref{coora}), respectively.

The position of satellites is suitably stated in the heliocentric-ecliptic coordinate system $(I,J,K)$. Its origin is at the Solar system heliocenter and $I-J$ plane coincides with the ecliptic plane. $I$ axis points along the direction of the vernal equinox, $K$ axis is perpendicular to the ecliptic plane, and $J$ axis is given by the right-handed coordinate system. The geocenter coordinates are given by the orbit of the Earth
\begin{eqnarray}\label{coore}
  I_{e}(t)&=&R_{e}\cos(\omega_{e}t+\theta_e)+\frac{R_{e}e_{e}}{2}\left( \cos 2(\omega_{e}t+\theta_e)-3\right)
 -\frac{3R_{e}e_{e}^{2}}{2}\cos(\omega_{e}t+\theta_e)\sin^{2}(\omega_{e}t+\theta_e),\nonumber\\
  J_{e}(t)&=&R_{e}\sin(\omega_{e}t+\theta_e)+\frac{R_{e}e_{e}}{2} \sin 2(\omega_{e}t+\theta_e)
   +\frac{R_{e}e_{e}^{2}}{4}\sin(\omega_{e}t+\theta_e)(3\cos 2(\omega_{e}t+\theta_e)-1),\nonumber\\
  K_{e}(t)&=&0,
\end{eqnarray}
where $R_{e}=1$ AU, $e_{e}=0.0167$ is the eccentricity of the Earth orbit, $\omega_{e}$ is the frequency of the Earth orbit, and $\theta_e$ represents the initial phase.

\begin{figure}
  \centering
  \includegraphics[width=0.4\textwidth]{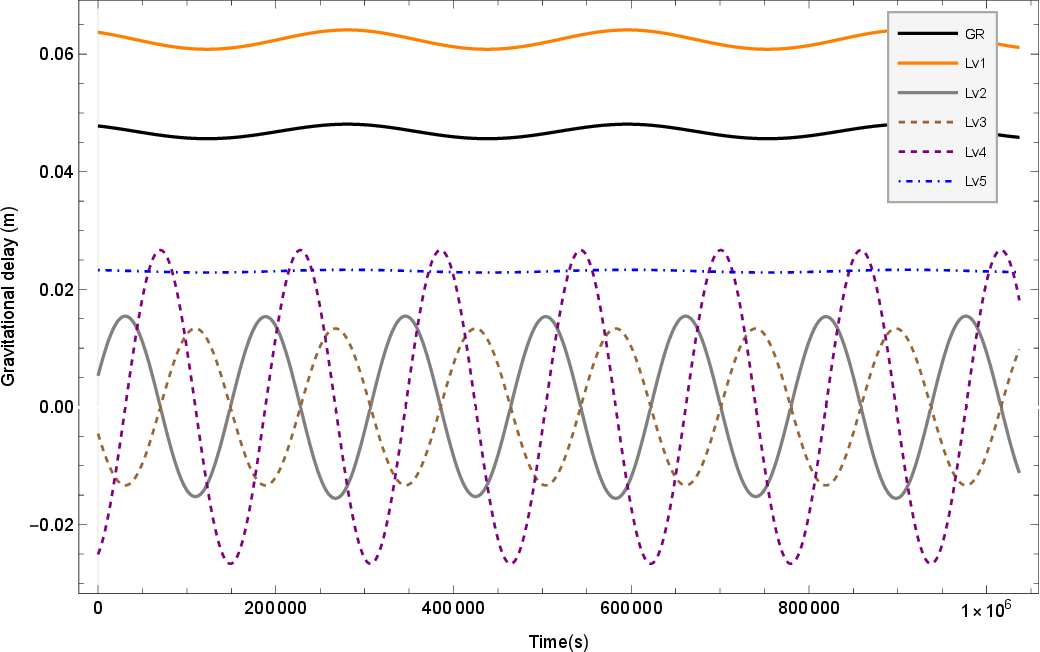}
  \caption{The gravitational delays due to the Earth contribution in the TianQin laser arms. The black curve is the standard GR contribution. The orange and gray curves represent the Lorentz-violating effects controlled by Lv1 and Lv2 terms, respectively. The brown and purple curves are the Lorentz-violating signals controlled by the coefficients combinations $\bar s^{11}-\bar s^{22}$ (labeled by Lv3) and $\bar s^{12}$ (labeled by Lv4), respectively. The blue curve is the signal contributed by the combination of Lorentz-violating coefficients $\bar s^{11}+\bar s^{22}-2 \bar{s}^{00}-2{\alpha}(\bar{a}^{E}_{\text{eff}})^{0}/{M_E}$ (labeled by Lv5). }\label{fig3}
\end{figure}

In the heliocentric-ecliptic coordinate system, the satellite $A$'s coordinates can be expressed as the superposition of the Earth's orbit $(I_e,J_e,K_e)$ and satellite orbital position $(x_{A},y_{A},z_{A})$ relative to the Earth. The satellite orbits are described by the orbital inclination $\beta_{s}$ and the longitude of the node $\phi_{s}$, which gives two corresponding rotations. Using the rotation matrices and coordinates (\ref{coora}), the heliocentric coordinates of satellite $A$ become
\begin{eqnarray}\label{cohel}
\left(
  \begin{array}{c}
    I_{A} \\
    J_{A} \\
    K_{A}
  \end{array}
  \right) &=&
  \left(
  \begin{array}{c}
    I_{e} \\
    J_{e} \\
    K_{e}
  \end{array}
  \right) +
   {\cal R}'\left({\phi_{s},\beta_{s}}\right)\left(
  \begin{array}{c}
     x_{A}\\
     y_{A} \\
     z_{A}
  \end{array}
  \right)
\end{eqnarray}
with
\begin{eqnarray*}\label{matrixos1}
{\cal R}' &=&
\left(
  \begin{array}{ccc}
    \cos\phi_s& -\sin\phi_s\cos\beta_s& \sin\phi_s\sin\beta_s \\
    \sin\phi_s &\cos\phi_s\cos\beta_s & -\cos\phi_s\sin\beta_s\\
    0 & \sin\beta_s & \cos\beta_s
  \end{array}
  \right).
\end{eqnarray*}
%where the angles are $\phi_{s}=30.5^\circ$ and $\beta_{s} = 94.7^\circ$.
The heliocentric coordinates for satellites $B$ and $C$ can be obtained from same method. Then, the satellite velocities can be calculated by
\begin{eqnarray}\label{cohev2}
\left(
  \begin{array}{c}
    \dot{I}_{b} \\
    \dot{J}_{b} \\
    \dot{K}_{b}
  \end{array}
  \right) &=&
  \left(
  \begin{array}{c}
    \dot{I}_{e} \\
    \dot{J}_{e} \\
    \dot{K}_{e}
  \end{array}
  \right) +
   {\cal R}'\left({\phi_{s},\beta_{s}}\right)\left(
  \begin{array}{c}
     \dot{x}_{b}\\
     \dot{y}_{b} \\
     \dot{z}_{b}
  \end{array}
  \right) ,
\end{eqnarray}
where the dot represents a time derivative with respect to coordinate time of heliocentric-ecliptic coordinate system, $b$ denotes satellites $A$, $B$, and $C$, and the first vector on the right-hand side of Eq.(\ref{cohev2}) is related to the velocity of the geocenter.

\begin{figure}
  \centering
  \includegraphics[width=0.4\textwidth]{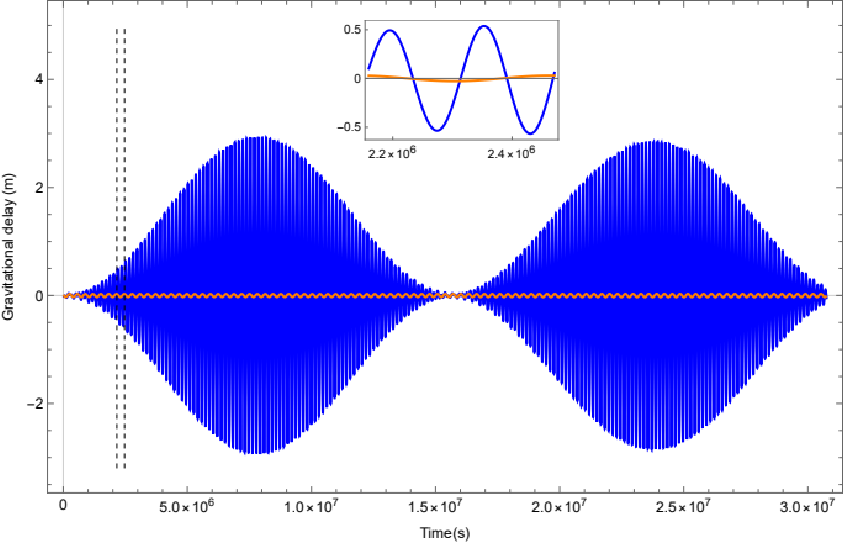}
  \caption{ The Lorentz-violating signals of the Sun gravitational delays in the interference signal. The orange curve is the signal of the first term in Eq.(\ref{gtimes2}). The blue cure represents the interference signals controlled by the secend term in Eq.(\ref{gtimes2}). The inset zooms in the region within the double dashed lines.}\label{fig4}
\end{figure}

Once the positions and velocities are given, it is easy to implement the gravitational delays, frequency shifts, and corresponding effects of the Lorentz violation.
To conveniently demonstrate the gravitational delays, we mark the first term of Eq.(\ref{gtime2}) as label Lv1, and mark the second term as label Lv2.
Figure.\ref{fig3} shows the Lorentz-violating contributions to gravitational delays induced by the Earth's gravitational field. The solid lines represent the contributions in the round-trip light path of one laser arm, the dashed lines indicate the contributions in the interference signal of adjacent laser arms, and the dotdashed line denotes the contribution in the closed-loop light path. The black curve is the standard GR contribution. The orange and gray curves represent the Lorentz-violating effects controlled by Lv1 and Lv2 terms, respectively (To plot, we set the value of the SME coefficients to be the order of 1). The brown and purple curves are the Lorentz-violating signals controlled by the coefficients combinations $\bar s^{11}-\bar s^{22}$ (labeled by Lv3) and $\bar s^{12}$ (labeled by Lv4), respectively. The blue curve is the signal contributed by the combination of Lorentz-violating coefficients $\bar s^{11}+\bar s^{22}-2\bar{s}^{00}-2{\alpha}(\bar{a}^{E}_{\text{eff}})^{0}/{M_E}$ (labeled by Lv5). Noted that the combinations of coefficients for Lorentz violation are sensitive to different frequencies or phases, which can be used to set experimental constraints.

\begin{figure}
  \centering
  \includegraphics[width=0.4\textwidth]{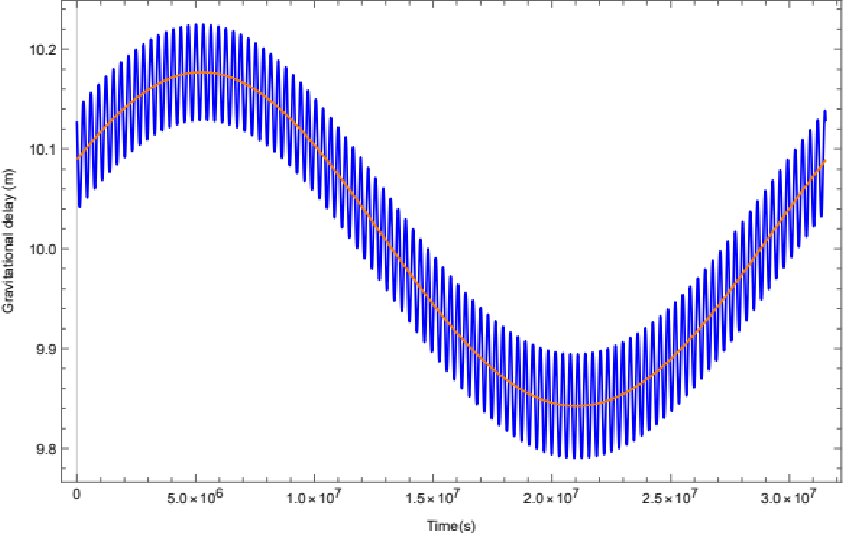}
  \caption{ The signals of the logarithmic term in Sun gravitational delays. The blue curve is the signal of single-arm round-trip path (it has been magnified 3 times for the purpose of comparison). The orange cure represents the gravitational-delay signals for the closed-loop light path.}\label{fig41}
\end{figure}

Figure.\ref{fig4} shows the Lorentz-violating signals of the Sun's gravitational delays in the interference signal. The orange curve is the of the logarithmic term in Eq.(\ref{gtimes2}), and the blue curve represents the Lorentz-violating signals controlled by the nonlogarithmic term of Eq.(\ref{gtimes2}). The inset presents the sinusoidal signals for Lorentz-violating effects within the dashed lines. The amplitude of the orange curve is the order of $10^{-2}$ m. The amplitude of the blue curve can reach 3 m and is modulated by the position of the GW detector with respect to the Sun. This motivates the experimental constraints on the time-component SME coefficient from the Sun gravitational delay effects.

Figure.\ref{fig41} shows the signals of the logarithmic term of Sun's gravitational delay. The blue curve is the signal of the round-trip of a single arm (for the purpose of comparison, we magnify this term 3 times), which includes annual variations and orbit-period variations. The orange cure represents the signals of the closed-loop light path, which contains only the annual variations. For the Sun's gravitational delays, the orbital-frequency variations cancel out in the closed-loop light path. In addition, there is the same characteristic for the gravitational-delay signals controlled by the coefficients of Lorentz violation in the nonlogarithmic term. This feature may help analyze for analyzing relativistic effects with data.

\begin{figure}
  \centering
  \includegraphics[width=0.4\textwidth]{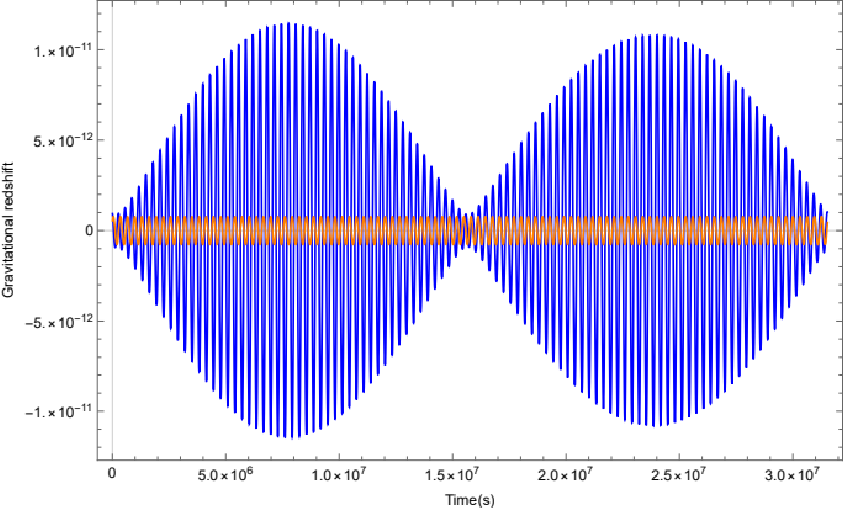}
  \caption{Gravitational redshifts. The blue and orange curves represent gravitational redshift signals of the Sun and Earth masses, respectively.}\label{fig5}
\end{figure}

\begin{figure}
  \centering
  \includegraphics[width=0.4\textwidth]{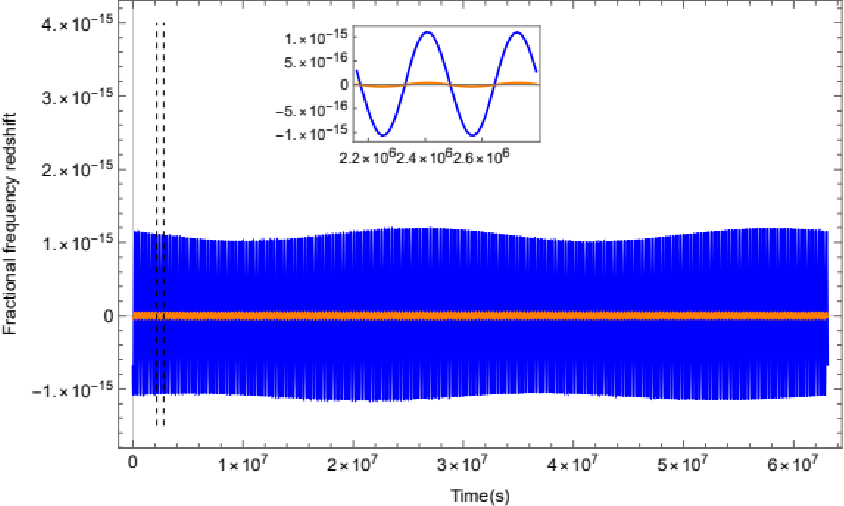}
  \caption{Gravitational fractional frequency shifts. The blue and orange curves represent gravitational fractional frequency redshift signals due to the Sun and Earth, respectively. The inset zooms in the region within the double dashed lines. }\label{fig6}
\end{figure}

We also evaluate the gravitational redshifts of the Sun and Earth masses, see Fig.\ref{fig5}. Owing to a small eccentricity, the Earth's gravitational redshift (orange curve) is suppressed to be small. Figure.\ref{fig6} shows the gravitational fractional frequency shifts of Eq.(\ref{freg}). The blue and orange curves represent the signals caused by the Sun and Earth's time delays, respectively. The Sun's and Earth's effects are the sinusoidal signals with amplitudes about $1\times10^{-15}$ and $4\times10^{-17}$, respectively. The amplitude for Sun's effect has a slight modulation due to the variations of the distance between the satellites configuration to the Sun.

\begin{figure}
  \centering
  \includegraphics[width=0.4\textwidth]{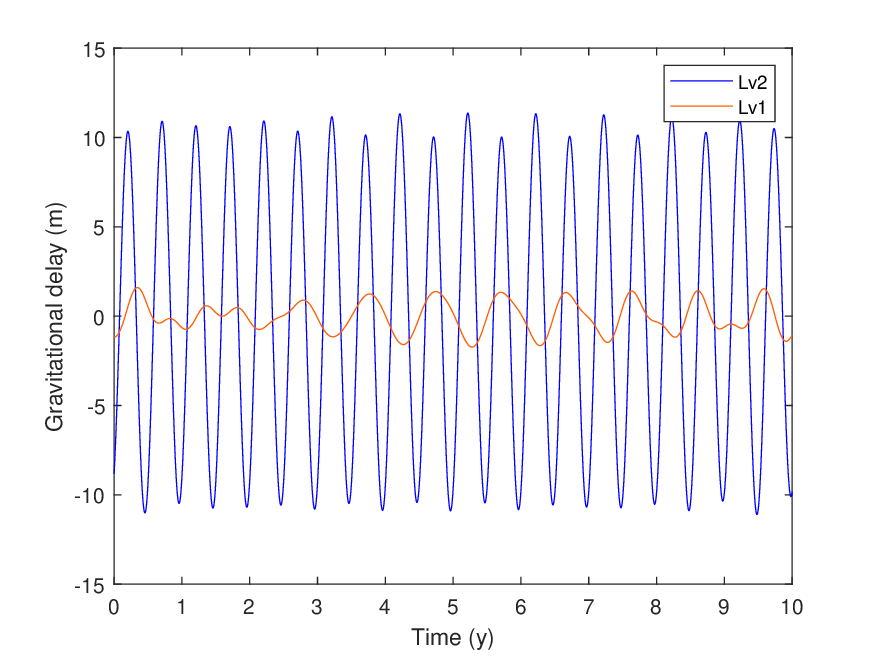}
  \caption{Gravitational delays of the Sun's contribution on the interference path of the LISA. The orange curve represents logarithmic-term effect, labeled as Lv1, and blue curve is nonlogarithmic-term effect, labeled by Lv2.}\label{fig71}
\end{figure}

\begin{figure}
  \centering
  \includegraphics[width=0.5\textwidth]{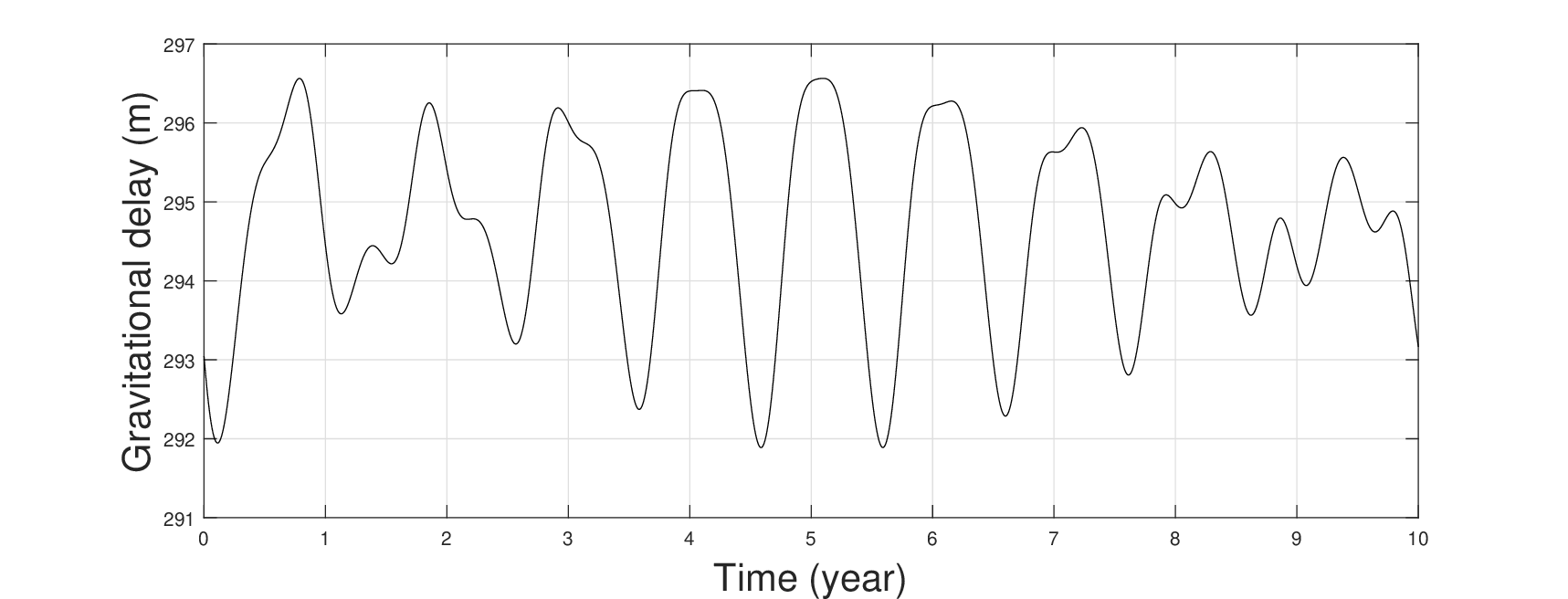}
  \includegraphics[width=0.5\textwidth]{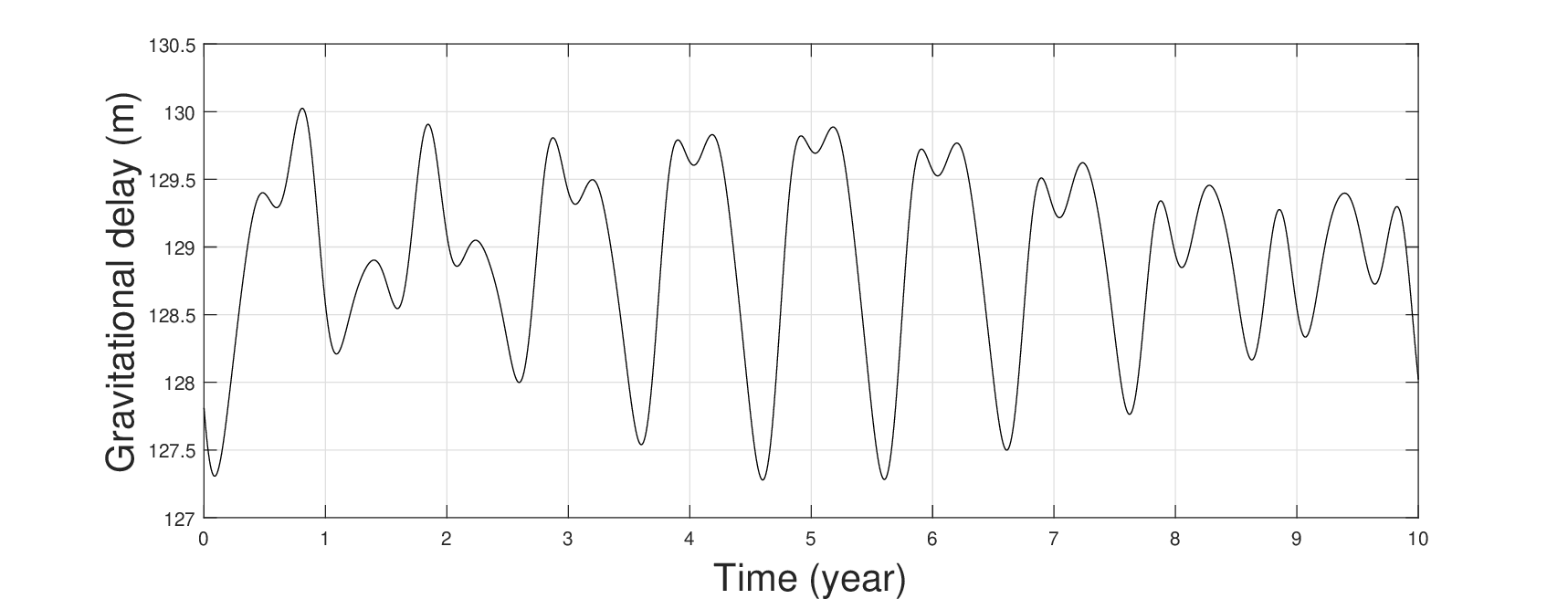}
  \caption{Gravitational delays of the Sun's contribution on the closed-loop path of the LISA. Upper panel: the contribution of logarithmic term. Lower panel: the contribution of nonlogarithmic term. }\label{fig72}
\end{figure}

\begin{figure}
  \centering
  \includegraphics[width=0.4\textwidth]{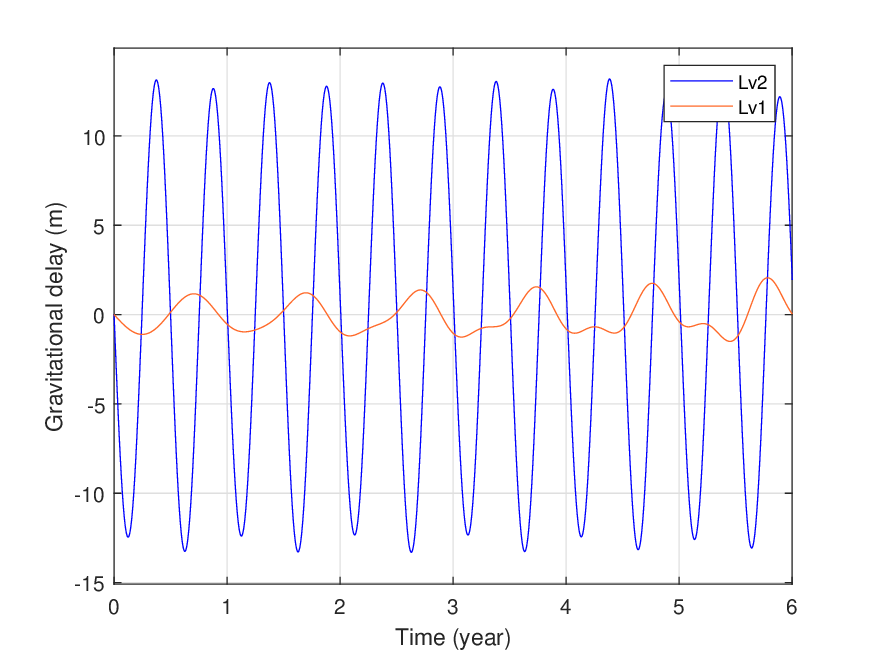}
  \caption{Gravitational delays of the Sun's contribution on the interference path of the Taiji. The orange curve represents the logarithmic-term effect, labeled as Lv1, and blue curve is nonlogarithmic-term effect, labeled by Lv2.}\label{fig81}
\end{figure}

\begin{figure}
  \centering
  \includegraphics[width=0.5\textwidth]{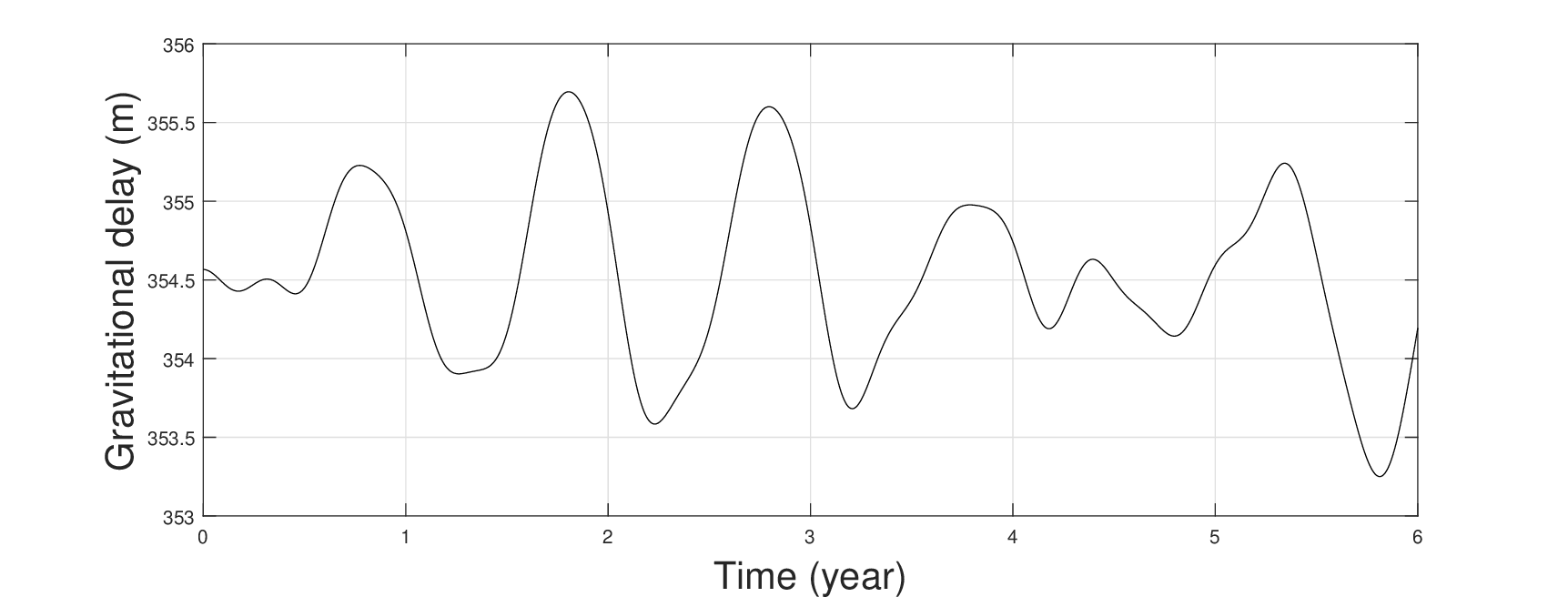}
  \includegraphics[width=0.5\textwidth]{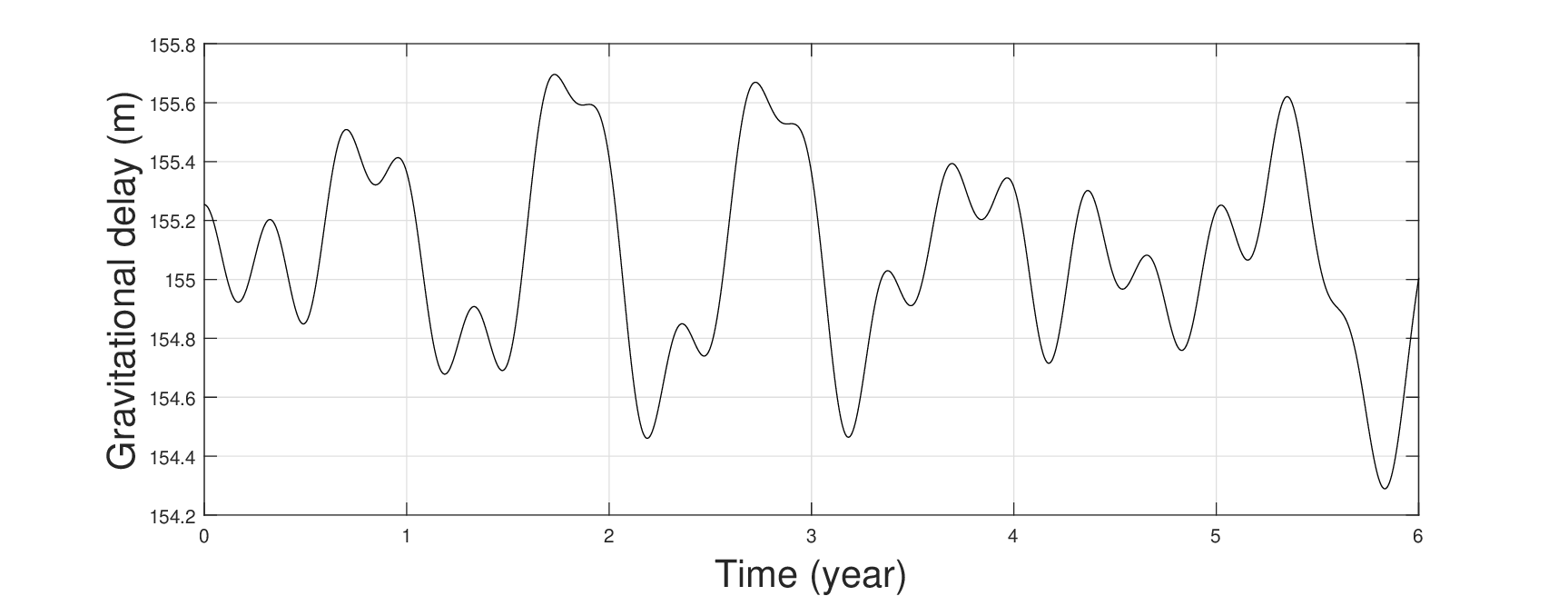}
  \caption{Gravitational delays of the Sun's contribution on the closed-loop path of the Taiji. Upper panel: the contribution of logarithmic term. Lower panel: the contribution of nonlogarithmic term. }\label{fig82}
\end{figure}

Considering the LISA and Taiji missions, the coordinates of satellites can be obtained by a similar process, and more accurate coordinates and velocities can be given by the high-precision orbit models \cite{martens2021trajectory,han2022effect}. Based on the LISA orbits \cite{martens2021trajectory}, it is convenient to evaluate the Lorentz-violating effects on data-streams combinations for the LISA mission. Figure. \ref{fig71} shows the Lorentz-violating contributions of the Sun's gravity. The orange and blue solid curves represent the logarithmic-term (label Lv1) and nonlogarithmic-term (label Lv2) effects on the interference path, respectively. In the interference signals, the logarithmic-term contributions are suppressed and nonlogarithmic-term contributions are enhanced to some extent. Figure. \ref{fig72}  present the logarithmic-term (upper panel) and nonlogarithmic-term (lower panel) contributions on the closed-loop path.
Based on the Taiji orbits \cite{han2022effect}, we calculated the response of Lorentz-violating effects to the combinations of Taiji data streams. Figure.\ref{fig81} show the logarithmic-term (label Lv1) and nonlogarithmic-term (label Lv2) Lorentz-violating contributions of the Sun's gravity in interference signals, respectively. Figure.\ref{fig82} plots the logarithmic-term (upper panel) and nonlogarithmic-term (lower panel) contributions on the closed-loop path, respectively.

\section{conclusion}\label{sct5}

In the context of the pure-gravity sector and matter-gravity coupling sector of the SME, we investigated the Lorentz-violating corrections to the gravitational time delays, and frequency shift of clocks. Based on these corrections, we analyze the influences of Earth's gravity and Sun's gravity on the space-based GW missions, TianQin, LISA, and Taiji.
For the gravitational time delays,  we proposed using the six data streams from the equilateral triangular configuration to construct the measurement-signal combinations, including the single-arm round-trip path, interference path, triangular round-trip path, and one-way closed-loop light path. These combinations are highly sensitive to various combinations of SME coefficients and provide independent linear combinations different previous studies. that can increase the diversity of experimental constraints. The formulas derived in our analysis can be used in the numerical model of GW missions. Based on the nominal concepts of TianQin, LISA, and Taiji, we estimated the attainable sensitivities for SME coefficients: $10^{-6}$ for the gravity sector coefficient $\bar{s}^{TT}$, $10^{-6}$ for matter-gravity coupling coefficients $(\bar{a}^{(e+p)}_{\text{eff}})_{T}$ and $\bar{c}^{(e+p)}_{TT}$, and $10^{-5}$ for $(\bar{a}^{n}_{\text{eff}})_{T}$ and $\bar{c}^{n}_{TT}$.
For the frequency shift, the GR and Lorentz-violating effects are determined up to the order of $c^{-3}$. We evaluated the influences of the gravitational redshift and gravitational frequency shift in space-based GW missions.

To illustrate our result, we evaluated the GR effects and corresponding Lorentz-violating effects for TianQin, LISA, and Taiji missions using their numerical orbits. By considering the positions and velocities of the satellites, we presented the signals of Lorentz violation and general relativity. Our numerical results showed that the combinations of SME coefficients are sensitive to signals with different frequencies or phases. Especially, for the equilateral triangular configuration, we observed the distinct characteristics in the gravitational signals from the Sun and Earth. The gravitational time delays present a unique opportunity to set constraints on time components of SME coefficients.

Lorentz symmetry is believed to be broken in various scenarios, such as unification, quantum gravity, and even some models of dark matter and dark energy. Although the absence of experimental evidence for Lorentz violation thus far, numerous opportunities for further study exist. There is still a large unexplored coefficients space that can be explored by improved measurement methods or by various projects. It is important to fully exploit existing experiments or future projects, such as space-based GW missions. These efforts will contribute to probing the vast parameters space as efficiently as possible and even searching for new physics.

\section{Acknowledgment}
The authors thank the anonymous referee for useful comments and constructive suggestions. This work is supported by the National Natural Science Foundation of China (Grants No.12247150, No.11925503 and No.12175076), and the Post-doctoral Science Foundation of China (Grant No.2022M721257).

\section{References}
\bibliographystyle{apsrev4-1}

\bibliography{tqsme}

\end{document}